\documentclass[aps,prl,showpacs,
amssymb,nofootinbib,twocolumn]{revtex4}

\usepackage{bm}
\usepackage{amsmath}
\usepackage{amssymb}
\usepackage{latexsym}
\usepackage{amsfonts}
\usepackage{epsfig}
\usepackage{psfrag}
\usepackage{graphicx}

\newcommand{\ket}[1]{\left|#1\right>}
\newcommand{\bra}[1]{\left<#1\right|}

\newcommand{\nn}{\nonumber\\}

\newcommand{\f}[1]{\mbox{\boldmath$#1$}}
\newcommand{\fk}[1]{\mbox{\boldmath$\scriptstyle#1$}}

\newcommand{\bea}{\begin{eqnarray}}
\newcommand{\ea}{\end{eqnarray}}
\newcommand{\eea}{\end{eqnarray}}
\newcommand{\ord}{{\cal O}}

\begin{document}

\title{Analogue of cosmological particle creation in an ion trap}

\author{Ralf Sch\"utzhold$^*$ and Michael Uhlmann}

\affiliation{Institut f\"ur Theoretische Physik,
Technische Universit\"at Dresden, D-01062 Dresden, Germany}

\author{Lutz Petersen, Hector Schmitz, Axel Friedenauer, 
and Tobias Sch\"atz$^\dagger$}

\affiliation{Max-Planck-Institut f\"ur Quantenoptik,
Hans-Kopfermann-Str.~1, D-85748 Garching, Germany}

\begin{abstract}
We study phonons in a dynamical chain of ions confined by a
trap with a time-dependent (axial) potential strength and demonstrate
that they behave in the same way as quantum fields in an
expanding/contracting universe.
Based on this analogy, we present a scheme for the detection of the
analogue of cosmological particle creation which should be feasible
with present-day technology.
In order to test the quantum nature of the particle creation mechanism 
and to distinguish it from classical effects such as heating, we propose 
to measure the two-phonon amplitude via the $2^{\rm nd}$ red side-band
and to compare it with the one-phonon amplitude
($1^{\rm st}$ red side-band).
\end{abstract}

 
\maketitle

{\em Introduction}\quad
%
The theory of quantum fields in curved space-times 
(see, e.g., \cite{birrell}) comprises many fascinating and striking
phenomena -- one of them being the creation of real particles out of
the (virtual) quantum vacuum fluctuations by a gravitational field.
These effects include Hawking radiation given off by black holes as 
well as cosmological particle creation.
A very similar mechanism -- the amplification of quantum vacuum
fluctuations due to the rapid expansion of the very early universe --
is (according to our standard model of cosmology) responsible for the
generation of the seeds for cosmic structure formation.  
Hence, even though these effects are far removed from every-day
experience,  they are very important for the past and the future fate 
of our universe. 

Therefore, it would be desirable to render these phenomena accessible
to an experimental verification.
Probably the most promising way for achieving this goal is to
construct a suitable analogue which reproduces the relevant features
(such as the Hamiltonian) of quantum fields in curved space-times.
Along this line of reasoning, proposals based on the analogy between
phonons in dynamical Bose-Einstein condensates and quantized fields in 
an expanding/contracting universe have been suggested
\cite{living}. 
Unfortunately, the detection of the created phonons in these systems
is rather difficult (see, however, \cite{detect}). 

On the other hand, the detection of single phonons in ion traps via
optical techniques is already state of the art in current
technology -- which suggests the study of this set-up instead. 
In this Letter, we shall derive the analogy between phonons in an
axially time-dependent ion trap and quantum fields in an
expanding/contracting universe and propose a corresponding detection
scheme for the analogue of cosmological particle creation.
A similar idea has already been pursued in \cite{alsing}, but the 
proposal presented therein goes along with several problems, which
will be discussed below \cite{kritik}. 

{\em Cosmological particle creation}\quad
%
Let us start by briefly reviewing the basic mechanism of particle
creation in an expanding/contracting universe. 
For simplicity, we consider a massless scalar field~$\phi$
described by the action, see, e.g., \cite{birrell}
($\hbar=c=1$ throughout)
\bea
\label{action}
{\cal A}=\frac12
\int d^4x\,\sqrt{|{\mathfrak g}|}
\left[
(\partial_\mu\phi){\mathfrak g}^{\mu\nu}(\partial_\nu\phi)
-\zeta{\mathfrak R}\phi^2
\right]
\,,
\ea
where ${\mathfrak g}^{\mu\nu}$ denotes the metric and 
${\mathfrak g}={\rm det}\{{\mathfrak g}_{\mu\nu}\}$ its determinant. 
Furthermore, a scalar field can be coupled to the Ricci (curvature)
scalar ${\mathfrak R}$ via a dimensionless parameter~$\zeta$
(e.g., conformal coupling $\zeta=1/6$, cf.~\cite{birrell}).
A spatially flat universe can be described in terms of the
Friedman-Robertson-Walker metric  
\bea
\label{metric}
ds^2={\mathfrak a}^6(t)dt^2-{\mathfrak a}^2(t)d\f{r}^2
\,,
\ea
with the time-dependent scale parameter ${\mathfrak a}(t)$
corresponding to the cosmic expansion/contraction.
Here we have chosen a slightly unusual time-coordinate $t$ related to
the proper time $\tau$ via $d\tau={\mathfrak a}^3(t)dt$ 
in order to simplify the subsequent formul\ae.
After a normal-mode expansion, the wave equation reads 
\bea
\label{normal-mode}
\ddot\phi_{\fk{k}}+
\left[
{\mathfrak a}^4(t)\f{k}^2+\zeta{\mathfrak a}^6(t){\mathfrak R}(t)
\right]
\phi_{\fk{k}}=0
\,,
\ea
i.e., each mode~$\f{k}$ just represents a harmonic oscillator with a
time-dependent potential 
${\mathfrak a}^4(t)\f{k}^2+\zeta{\mathfrak a}^6(t){\mathfrak R}(t)$. 
As long as this external time-dependence of the potential is much
slower than the internal frequency of the oscillator, 
the quantum state will stay near the ground state due to the adiabatic
theorem. 
However, if the external time-dependence is fast enough 
(i.e., non-adiabatic), the evolution will transform the initial ground
state into an excited state in general -- which is the basic mechanism
of cosmological particle creation.
In this case, the initial vacuum state 
$\ket{0}=\ket{\psi(t\downarrow-\infty)}$ containing no particles 
$\forall_{\fk{k}}\,\hat a_{\fk{k}}\ket{0}=0$ evolves into a squeezed
state    
\bea
\label{squeezed}
\ket{\psi(t\uparrow\infty)}
&=&
\exp\left\{\sum\limits_{\fk{k}}
\xi_k\,\hat a_{\fk{k}}^\dagger\,\hat a_{-\fk{k}}^\dagger
-{\rm h.c.}
\right\}\ket{0}
\nn
&=&
\ket{0}+\sum\limits_{\fk{k}}
\xi_k\ket{1_{\fk{k}},1_{-\fk{k}}}+\ord(\xi_k^2)
\,,
\ea
which does contain pairs of particles $\ket{1_{\fk{k}},1_{-\fk{k}}}$. 
The squeezing parameter $\xi_k$ for each mode is determined by
the solution of Eq.~(\ref{normal-mode}) and thus by the
time-dependence of ${\mathfrak a}^4(t)\f{k}^2$ 
as well as $\zeta{\mathfrak a}^6(t){\mathfrak R}(t)$ 
and governs the number of created particles per mode 
\bea
\label{number}
\langle\hat n_{\fk{k}}\rangle
=
\bra{\psi(t\uparrow\infty)}
\hat a^\dagger_{\fk{k}}\hat a_{\fk{k}}
\ket{\psi(t\uparrow\infty)}
=
\sinh^2(|\xi_k|)
\,.
\ea
%

{\em Ion-trap analogue}\quad
%
Assuming a strong radial confinement of the ions, we consider their 
axial motion only.
In a time-dependent harmonic axial potential described by the
oscillator frequency $\omega_{\rm ax}(t)$, the position~$q_i$ 
of the $i$-th ion obeys the equation of motion 
\bea
\label{eom-full}
\ddot q_i+\omega_{\rm ax}^2(t)q_i
=
\gamma\sum\limits_{j \neq i}
\frac{{\rm sign}(i-j)}{(q_i-q_j)^2}
\,,
\ea
where the factor $\gamma$ encodes the strength of the Coulomb
repulsion between the ions. 
Assuming a static situation initially, the classical solution to the
above equation can be obtained via the scaling ansatz 
$q_i(t)=b(t)q_i^0$, where $q_i^0$ are the initial static equilibrium
positions, leading to the evolution equation for the scale parameter
$b(t)$ 
\bea
\label{scaling}
\ddot b+\omega_{\rm ax}^2(t)b
=
\frac{\omega_{\rm ax}^2(0)}{b^2}
\,.
\ea
In order to treat the quantum fluctuations of the ions 
(leading to the quantized phonon modes), let us split the full
position operator $\hat q_i(t)$ for each ion into its classical
trajectory $b(t)q_i^0$ and quantum fluctuations $\delta\hat q_i(t)$
\bea
\label{split}
\hat q_i(t)=b(t)q_i^0+\delta\hat q_i(t)
\,.
\ea
Since these fluctuations $\delta\hat q_i(t)$ are very small for heavy
ions, we may linearize the full equation of motion~(\ref{eom-full})
\bea
\label{linearization}
\left(\frac{\partial^2}{\partial t^2}+\omega_{\rm ax}^2(t)\right)
\delta\hat q_i
=
\frac{1}{b^3(t)}
\sum\limits_{j}
M_{ij}\delta\hat q_j
\,,
\ea
with a time-independent matrix $M_{ij}$ arising from the Coulomb
term in~(\ref{eom-full}).
Diagonalization of this matrix (normal-mode expansion)
yields the phonon modes
\bea
\label{kappa}
\left(
\frac{\partial^2}{\partial t^2}+\omega_{\rm ax}^2(t)
+\frac{\omega_\kappa^2}{b^3(t)}
\right)
\delta\hat q_\kappa
=0
\,,
\ea
labeled by $\kappa$.  
The time-independent eigenvalues $\omega_\kappa^2\geq0$ of the
matrix $M_{ij}$ determine the phonon frequencies.
The lowest mode is the center-of-mass mode corresponding to a
simultaneous (rigid) motion of the ions.
Since the ion distances are fixed, the Coulomb term does not
contribute in this situation $\omega_\kappa^2=0$.
The next mode is the breathing mode with 
$\omega_\kappa^2=2\omega_{\rm ax}^2(0)$. 
Comparing Eqs.~(\ref{scaling}) and (\ref{kappa}), we see that this
mode exactly corresponds to the scaling ansatz, i.e., the ion cloud
expands/contracts linearly.
Hence this is the only mode which can be excited classically 
(for a purely harmonic potential).
I.e., without imperfections such as heating, phonons in the other
modes can only be created by quantum effects. 

Comparing Eqs.~(\ref{normal-mode}) and (\ref{kappa}) and identifying 
$\phi_{\fk{k}}$ with $\delta\hat q_\kappa$, we observe a strong
similarity:
The wavenumber $\f{k}^2$ in ~(\ref{normal-mode})
directly corresponds to $\omega_\kappa^2$ in~(\ref{kappa}) 
and the scale factors ${\mathfrak a}(t)$ and $b(t)$ enter in a similar
way. 
However, an expanding universe is analogous to a contracting ion
cloud and vice versa.
In the mode-independent terms, the axial trap 
frequency~$\omega_{\rm ax}$ acts like the Ricci 
scalar~${\mathfrak R}$. 
Interestingly, both are related to the second time-derivatives of the  
corresponding scale factors.

In view of the formal equivalence of Eqs.~(\ref{scaling}) and
(\ref{kappa}), we obtain the same effects as in cosmology -- 
in particular, the mixing of creation and annihilation operators 
\bea
\label{Bogolubov}
\hat a_\kappa(t\uparrow\infty)
=
\alpha_\kappa\hat a_\kappa(0)
+\beta_\kappa\hat a_\kappa^\dagger(0)
\,,
\ea
which can be expressed in terms of the Bogolubov coefficients
satisfying  
$|\alpha_\kappa^2|-|\beta_\kappa^2|=1$.
Note that the above relation is just the operator representation of
the squeezing transformation in Eq.~(\ref{squeezed}) with 
$|\beta_\kappa|\leftrightarrow\sinh(|\xi_k|)$. 

\begin{figure}
\vspace*{-0.5cm} 
\begin{center}
\includegraphics*[width=\columnwidth]{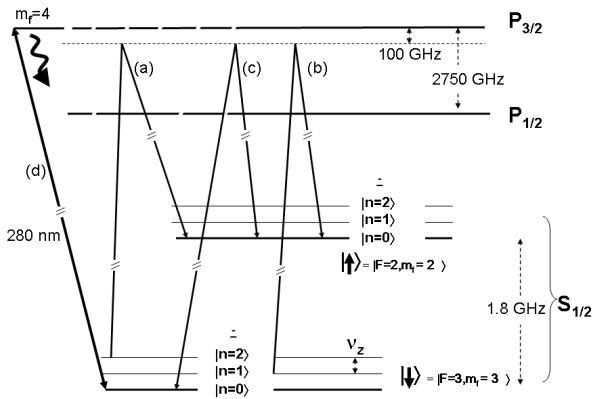}
\end{center}
\vspace*{-0.5cm} 
\caption{Schematic of the relevant energy levels (not to scale) of one 
$^{25}$Mg$^+$ ion. 
Shown are the ground-state hyperfine levels supplying the two internal 
states ($\ket{\downarrow}$ and $\ket{\uparrow}$) and the 
first three equidistant harmonic oscillator levels $\ket{n}$, related 
to the harmonic axial confinement in a linear ion trap. 
Typically, the energy splitting of the motional levels and the Zeeman 
shift induced by an external magnetic field are of the same order of 
magnitude within 1-10 MHz, therefore much smaller than the Hyperfine 
splitting of 1.8 GHz, the fine structure splitting of 2750 GHz and the 
optical transition frequency of the order of $10^{15}$Hz. 
We depict the resonant transition state sensitive detection named (d) 
and the relevant types of off- resonant ($\approx$ 100 GHz) two-photon
stimulated Raman transitions (a,b and c) described in the text.}
\label{Figure:1}
\end{figure}

\begin{figure}
\begin{center}
\includegraphics*[width=\columnwidth]{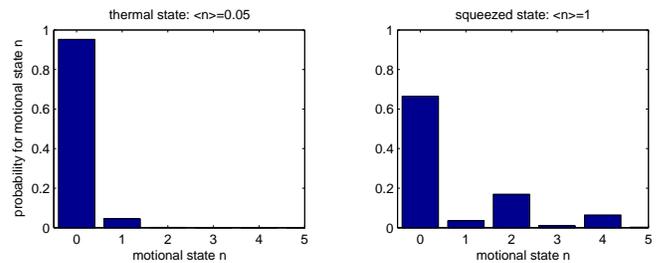}
\end{center}
\vspace*{-0.5 cm} 
\caption{Population of the motional states: On the
left hand side, we show the experimentally realized thermal spectrum
after cooling the system close to the ground state 
($\langle\hat n\rangle$=0.05). On the right hand side, we show the simulated
spectrum of the ($\langle\hat n\rangle$=1) squeezed state, assuming
the starting distribution from the left. The almost complete absence
of population in odd motional states emphasizes the non-classical
character of this state and guides towards a simple possibility to
distinguish it from thermal and disturbed ones via measuring a much
higher population in state $|n=2\rangle$ than $|n=1\rangle$.}
\label{Figure:2}
\end{figure}

{\em Detection scheme}\quad
%
In the following we describe how to realize the proposed experiment
by applying operations closely related to those implemented on
qubit-ions in  quantum information processing~\cite{bible98}. 
We focus on initializing the system, simulating the non-adiabatic 
expansion of space, performing the read-out of the final state 
(particle- or phonon-number distribution) and distinguishing it 
from a classically describable outcome, for example caused by 
thermal heating.
To perform a first realization, we will confine one single earth
alkaline atomic ion to the axis of a linear radio-frequency
trap~\cite{pqe07} similar to that described in~\cite{rowe02,ferdi03}. 
The required simulation basis can be composed by a
$^2$S$_{1/2}$ electronic ground state level of $^{25}$Mg$^+$, here
the state $|F=3;m_f= 3\rangle=\ket{\downarrow}$, and the
associated harmonic oscillator levels $|n\rangle$ related to the
axial harmonic confinement, as depicted in Fig.~\ref{Figure:1}. 
At the start of each experiment, the ion will be laser cooled close
to the ground state of the axial (external) motion and optically
pumped into the electronic (internal) state
$\ket{\downarrow}$~\cite{initial98}.
Then we will decrease adiabatically the axial confinement and
subsequently reset it non-adiabatically to its initial value 
(as already proposed in~\cite{heinz90} in another context). 
Since the ground state wave function of the ion cannot adapt to the 
restored stiff confinement (non-adiabatic case), it will oscillate 
symmetrically around the minimum of the final trapping 
potential~\cite{didihabil03}, 
i.e., without populating odd motional states. 
As shown above, this non-classical oscillation is to be described 
via a squeezed state (see also \cite{motion96}) 
depicted in Fig.~\ref{Figure:2}.

We propose, in addition to established schemes described
in~\cite{motion96} or~\cite{enrique05}, for example, an alternative 
method to distinguish classical noise (such as the initial thermal 
distribution or heating during the process) from a squeezed state 
generated by quantum effects considered here.
In order to read out the final motional state, we will first couple 
it (via suitable lasers) to two internal states of the ion. 
Besides the electronic ground state 
$|F=3;m_f= 3\rangle=\ket{\downarrow}$, the second internal state 
to span a two-level system (analogous to a qubit) is implemented 
via a second hyperfine state of $^{25}$Mg$^+$, 
$|F=2,m_f=2\rangle \equiv \ket{\uparrow}$, separated from
the state $\ket{\downarrow}$ by the hyperfine splitting
$\omega_\mathrm{o} \simeq 2\pi \times 1.8$ GHz. 
We will accomplish the coupling of the two internal states 
$\ket{\downarrow}$, $\ket{\uparrow}$ and the motional 
states $|n\rangle$ via two-photon stimulated Raman 
transitions~\cite{bible98} requiring two laser beams 
($\lambda \approx 280 $ nm), with wave vector difference
$\Delta\f{k} = \f{k_2}-\f{k_1}$ aligned along the trap
axis $z$ ($|\Delta k|= \sqrt{2} \times 2\pi/\lambda =
2\pi/\lambda_\mathrm{eff}$). 
Via detuning the frequency difference $\omega_2 - \omega_1$ from 
the hyperfine splitting $\omega_\mathrm{o} \pm 2 \pi m \nu_z$
by integer multiples $m$ of the axial trapping frequency $\nu_z$,
we may drive the carrier-transition ($m=0$) or the first- ($m=1$) 
and second-($m=2$) sideband transitions respectively.
Note that the spectral resolution of the two-photon stimulated Raman
transition is independent on the natural line width 
$\Gamma = 2 \pi\times$43~MHz of the resonant transition -- but 
proportional to the inverse of the Rabi-frequency, adjusted via the 
intensities or detuning of the laser beams allowing for the resolution 
of the individual motional states separated by $\nu_z \ll \Gamma $.
In order to measure the population of the motional state $|n=2\rangle$,
we will drive a sequence of transitions (cf.~Fig.~\ref{Figure:1}), 
synthesized by a second-sideband (a) transition 
($\ket{\downarrow,n=2}\rightarrow \ket{\uparrow,n=0}$) 
followed by a carrier (c) transition 
($\ket{\uparrow,n=0} \rightarrow \ket{\downarrow,n=0}$).
The final read-out (d) described below is internal-state dependant 
and provides us with the population in state $\ket{\downarrow}$.
After the sequence (a,c,d) of transitions this is almost exclusively
equivalent to the population of the motional state $|n=2\rangle$ 
(because the probability of even higher excitations $|n\geq3\rangle$
is expected to be much smaller and their Rabi frequencies are also
different). 
This result can then be compared with the outcome after a first 
red-sideband (b) transition 
($\ket{\downarrow,n=1} \rightarrow \ket{\uparrow,n=0}$) 
followed by a carrier (c) transition
($\ket{\uparrow,n=0} \rightarrow \ket{\downarrow,n=0}$), 
providing the probability of motional excitation $|n=1\rangle$. 
As soon as we deduce a higher probability for motional state 
$|n=2\rangle$ than for motional state $|n=1\rangle$, we show the 
incompatibility with classical effects such as a thermal distribution 
and get strong evidence for the non-classical effect of
squeezing~\cite{cohstate}.

Finally, we have to read out the internal state efficiently. 
To this end, we apply an additional resonant laser beam (d), 
tuned to a cycling transition~\cite{bible98}, coupling only state 
$\ket{\downarrow}$ resonantly to the $P_{3/2}$
level and providing spontaneous emission at rates of $>$ 10 MHz.
This allows to distinguish the ``bright'' 
$\ket{\downarrow}$ from the ``dark'' 
$\ket{\uparrow}$ state with high accuracy,
even at a low detection efficiency (due to the restricted
solid angle etc.). 

{\em Envisioned results}\quad
%
The above mentioned sate of the art techniques allow to cool the
axial motion close to the ground state 
$\langle\hat n\rangle\approx0.05$~\cite{pqe07} 
(see also \cite{initial98,ferdi03}) 
and to optically pump into the down state $\ket{\downarrow}$
with 99$\%$~\cite{pqe07} or even higher
fidelity~\cite{initial98,ferdi03}. 
First experiments show a possible non-adiabatic variation of the
axial motional frequency $\nu_z$ between 200 kHz and $\ge$ 2 MHz
with a related rise time of the order of one micro-second, which  
is sufficiently fast compared to the oscillation period of the lower 
frequency. 
Numerical simulations (based on measured temporal variation curves)
indicate that we should be able to transfer approximately 20$\%$ 
of the motional state population from the ground state $|n=0\rangle$ 
into state $|n=2\rangle$, which corresponds to a squeezed state with 
$\langle\hat n\rangle\approx1$.
Starting with a thermal distribution with 
$\langle\hat n\rangle\approx0.05$ instead
of the exact ground state $|n=0\rangle$, there will also be a small
final population (a few percent) of the state $|n=1\rangle$, 
see Fig.~\ref{Figure:2}.
However, this residual effect will be significantly smaller than 
the $|n=2\rangle$ population such that the signatures of squeezing
can be measured as described above. 

{\em Conclusions}\quad
%
Since the state of the art fidelities for the carrier and sideband 
transitions as well as the state sensitive detection exceed 
99$\%$~\cite{initial98,ferdi03}, the initialization and
measurement of the system can be provided with high accuracy. 
In order to benefit from these operational fidelities, the main task 
will be to minimize classical disturbances.  
For example, we have to carefully balance the applied voltages for
confinement during their non-adiabatic changes to prevent classical
excitation of the axial motional mode. 
In comparison to some other experiments with ion traps, the requirements 
for the present proposal may be a bit easier to achieve because the 
duration of the experiment will be short (around 3 ms) compared to the 
inverse of the thermal heating rate for motional quanta inside the trap 
($\leq$0.005 quanta/ms~\cite{pqe07}) and because the thermal and the
squeezed motional spectra show fundamentally different
characteristics, see Fig.~\ref{Figure:2}.
It should also be emphasized that it is impossible to resolve
individual motional states with pulse durations short compared to the
inverse of their frequency difference 
(see also \cite{alsing} and \cite{kritik}). 
This impossibility in resolution is related to the Heisenberg
uncertainty principle $\Delta E\Delta t\geq\hbar/2$ that allows to
create pairs of particles (phonons) out of the vacuum (ground) state
in first place. 
Increasing the system towards 8 modes (ions) might be feasible by
this proposal with state of the art techniques~\cite{natures05},
further scaling might benefit from the technical progress driven by
the attempts of the quantum information community.

Apart from experimentally testing the analogue of cosmic particle
creation -- which might ultimately allow the study of the impact of  
decoherence and interactions etc. --  
the investigation of non-adiabatic switching of trapping potentials
and its influence on the quantum state on motion might also shed
light on possible problems in schemes where the fast shuttling of
ions in a multiplex trap architecture is required for scaling
towards a universal quantum computer.

\acknowledgments
%
{\em Acknowledgments}\quad
%
This work was supported by the Emmy-Noether Programme of the
German Research Foundation (DFG, grants SCHU~1557/1-2 and
SCHA~973/1-2) and partly by the MPQ Garching. 

$^*$\,{\scriptsize\sf schuetz@theory.phy.tu-dresden.de} 
$^\dagger$\,{\scriptsize\sf  Tobias.Schaetz@mpq.mpg.de} 



\begin{thebibliography}{9999}
\bibitem{cohstate}
To exclude the possibility of a coherently driven state (and the
resulting higher excitations), we would detect the population of the
motional state $|n=0\rangle$ via a another sideband transition 
($\ket{\downarrow,n}\to\ket{\uparrow,n-1}$) which is not able to
transfer population out of the state $\ket{\downarrow,n=0}$.
\end{thebibliography}
\end{document}